\newskip\humongous \humongous=0pt plus 1000pt minus 1000pt

\newif\ifdtup

\def\beq{\begin{equation}}
\def\eeq{\end{equation}}

\def\beqn{\begin{eqnarray}}
\def\eeqn{\end{eqnarray}}
\relax

\def\dotx{\dotx{\dot\overline{x}}}

\catcode`\@=11

\@addtoreset{equation}{section}

\def\@normalsize{\@setsize\normalsize{15pt}\xiipt\@xiipt
\abovedisplayskip 14pt plus3pt minus3pt%
\belowdisplayskip \abovedisplayskip
\abovedisplayshortskip  \z@ plus3pt%
\belowdisplayshortskip  7pt plus3.5pt minus0pt}

\def\small{\@setsize\small{13.6pt}\xipt\@xipt
\abovedisplayskip 13pt plus3pt minus3pt%
\belowdisplayskip \abovedisplayskip
\abovedisplayshortskip  \z@ plus3pt%
\belowdisplayshortskip  7pt plus3.5pt minus0pt

\def\@listi{\parsep 4.5pt plus 2pt minus 1pt
            \itemsep \parsep
            \topsep 9pt plus 3pt minus 3pt}}

\def\underline#1{\relax\ifmmode\@@underline#1\else
        $\@@underline{\hbox{#1}}$\relax\fi}
\@twosidetrue

\catcode`@=12

\catcode`\@=11

\catcode`\@=12

\relax

\def\figcap{\section*{Figure Captions\markboth
        {FIGURECAPTIONS}{FIGURECAPTIONS}}\list
        {Fig. \arabic{enumi}:\hfill}{\settowidth\labelwidth{Fig. 999:}
        \leftmargin\labelwidth
        \advance\leftmargin\labelsep\usecounter{enumi}}}
 \relax
\def\tablecap{\section*{Table Captions\markboth
        {TABLECAPTIONS}{TABLECAPTIONS}}\list
        {Table \arabic{enumi}:\hfill}{\settowidth\labelwidth{Table 999:}
        \leftmargin\labelwidth
        \advance\leftmargin\labelsep\usecounter{enumi}}}
 \relax
\def\reflist{\subsubsection*{References\markboth
        {REFLIST}{REFLIST}}\list
        {[\arabic{enumi}]\hfill}{\settowidth\labelwidth{[999]}
        \leftmargin\labelwidth
        \advance\leftmargin\labelsep\usecounter{enumi}}}
 \relax

\catcode`\@=11

\def\FERMIPUB{}

\def\ps@headings{\def\@oddfoot{}\def\@evenfoot{}
\def\@oddhead{\hbox{}\hfill
        \makebox[.5\textwidth]{\raggedright\ignorespaces --\thepage{}--
        \hfill {\rm FERMILAB--Pub--\FERMIPUB}}}
\def\@evenhead{\@oddhead}
\def\subsectionmark##1{\markboth{##1}{}}
}

\ps@headings

\relax

\input space2st
\input math_mac
\documentstyle[art12,A4]{article}
\begin{document}
\begin{titlepage}
\begin{flushright}
Z\"urich University ZU-TH 31/93\\
\end{flushright}
\vfill
\begin{center}
{\large\bf ON THE MASS OF THE DARK COMPACT HALO OBJECTS}\\
\vfill
{\bf Ph.~Jetzer$^1$* and E.~Mass\'o$^2$}\\
\vskip 1.0cm
$^1$Institute of Theoretical Physics, University of Z\"urich,
Winterthurerstrasse 190,\\
CH-8057 Z\"urich, Switzerland\\
$^2$Departament de F\' \i sica and IFAE, Universitat Aut\`onoma
de Barcelona,\\
 E-08193 Bellaterra, Spain.
\end{center}
\vfill
\begin{center}
Abstract
\end{center}
\begin{quote}
Recently the French EROS collaboration and the American-Australian
MACHO collaboration have reported the observation of altogether
three possible microlensing events by monitoring over several
years the brightness of millions of stars in the Large Magellanic
Cloud. For each of these events, assuming they are due to microlensing,
we compute the most likely mass for the dark compact halo object, which acted
as gravitational lens. The most likely masses are 0.12, 0.31 and
0.38 $M_{\odot}$.
The average mass calculated using the method of moments
turns out to be 0.14 $M_{\odot}$.
\end{quote}
\vfill
\begin{flushleft}
\vfill
\begin{center}
December 1993
\end{center}
\vskip 0.5cm
$^*$ Supported by the Swiss National Science Foundation.
\end{flushleft}
\end{titlepage}
\newpage

One of the most important problems in astrophysics is the nature of the dark
matter present in the galactic halos, which is inferred from the
shape of the measured rotation curves. A possible candidate for the dark
halos are ``brown dwarfs'' or Jupiter like bodies
which are aggregates of the primordial elements: Hydrogen and Helium,
and have masses in the range ${\rm O}(10^{-7}) < M/M_{\odot} <
{\rm O} (10^{-1})$  \cite{kn:Derujula1}.
Paczy\'nski suggested a way to detect such objects in the halo of our
own galaxy using the gravitational lens effect \cite{kn:Pac}.

Recently the French collaboration EROS \cite{kn:EROS}
and the American-Australian
collaboration MACHO \cite{kn:MACHO} reported the possible detection of
altogether three microlensing events, discovered by monitoring over several
years millions of stars in the Large Magellanic Cloud (LMC).
Another event has been reported by a
Polish-American collaboration \cite{kn:Udalski},
which however monitors the galactic bulge.
Assuming that these observations are true microlensing events
we compute, using the formalism developed in ref.\cite{kn:Derujula},
for each event
the most likely mass for the dark halo object, which acted as
gravitational lens, as well as the average mass using the method of
moments. Of course the very small number of events does not
allow a precise mass distribution determination, so that
our results are preliminary.
Nevertheless, it shows how in practice it will
be possible to get precise information, as soon as
a sufficient number of microlensing events will be available, on the mass
distribution as well
as to what fraction they contribute to the total
dark halo mass.
In the following we use the formulas and notation derived
in ref.\cite{kn:Derujula} to which we refer for a detailed
explanation.

First we compute the probability $P$ that a microlensing
event of duration T
and maximum amplification $A_{max}$ be produced by a massive halo object (MHO)
of mass $\mu$ ($\mu$ is the mass of the MHO in Jupiter mass units,
$M_J = 0.95 \times 10^{-3} M_{\odot}$).
Let $d$ be the distance of the MHO from the line of
sight between the observer and a star in the LMC, t=0 the instant
of closest approach, and $v_T$ the MHO velocity in the transverse
plane. The magnification $A$ as a function of time is calculated using
simple geometry, and is given by
\begin{equation}
A(t)=A[u(t)]=\frac{u^2+2}{u(u^2+4)^{1/2}}~, \label{eqno:1}
\end{equation}
where
\begin{equation}
u^2=\frac{d^2+v_T^2 t^2}{R_E^2}~. \label{eqno:2}
\end{equation}
The light curve is a universal function determined by the two
parameters: $d/R_E$ and $v_T/R_E$. $R_E$ is the Einstein radius
which is
\begin{equation}
R_E^2=\frac{4GMD}{c^2}x(1-x)=r_E^2 \mu x(1-x)  \label{eqno:3}
\end{equation}
with $M$ (respectively $\mu$) the MHO mass (in $M_J$ Jupiter mass units)
and $D~ (xD)$ the distance
from the observer to the source (to the MHO). $D = 55~ kpc$ is the distance
to the LMC, and $r_E=9.78 \times 10^7~ km$.
We use here the definition: $T=R_E/v_T$ (this is slightly
different from the definition used in ref.\cite{kn:Derujula}).

We adopt the model of an isothermal spherical halo in which the
normalized MHO number distribution as a function of $v_T$ is
\begin{equation}
f(v_T) dv_T=\frac{2}{v_H^2} v_T e^{-v_T^2/v^2_H} dv_T~ , \label{eqno:4}
\end{equation}
with $v_H \approx 210~ km/s$ the velocity dispersion implied by the
rotation curve of our galaxy.
The MHO number density distribution per unit mass $dn/d\mu$ is
given by
\begin{equation}
\frac{dn}{d\mu}=H(x)\frac{dn_0}{d\mu}=\frac{a^2+R^2_{GC}}
{a^2+R^2_{GC}+D^2x^2-2DR_{GC} x cos \alpha}~\frac{dn_0}{d\mu}, \label{eqno:5}
\end{equation}
with $dn_0/d\mu$ the local MHO mass distribution.
We have assumed that $dn/d\mu$ factorizes in functions of $\mu$ and $x$
\cite{kn:Derujula}.
We take $a = 5.6~ kpc$ the galactic ``core'' radius
(our final results do not depend much on the poorly known value of $a$),
$R_{GC} = 8.5~ kpc$
our distance from the centre of the galaxy, and $\alpha = 82^0$
the angle between the line of sight and the direction of the galactic
centre.
For an experiment monitoring $N_{\star}$ stars during a total observation
time $t_{obs}$ the number of expected microlensing events is
given by \cite{kn:Derujula,kn:Griest}
\begin{equation}
N_{ev}=N_{\star} t_{obs}
2Dr_E \int v_T f(v_T) (\mu x(1-x))^{1/2} H(x) \frac{dn_0}{d\mu}
d\mu du_{min} dv_T dx   \label{eqno:6}
\end{equation}
where the integration variable $u_{min}$ is related to $A_{max}$:
$A_{max}=A[u = u_{min}]$.
The integration ranges being: $0 \leq x \leq 1$; $0 \leq v_T \leq 3v_T$, but
the upper (escape-velocity) limit can be substituted by
infinity with insignificant effect on the result;
$0 \leq u_{min} \leq u_{TH}$
(the minimal experimentally detectable magnification $A_{TH}$ is related
to $u_{TH}$: $A_{TH}=A[u = u_{TH}]$; and finally $0 \leq \mu \leq
\infty$ (for a more complete discussion see ref.\cite{kn:Derujula}).

In order to get the probability distribution
$P$ we perform the following variable transformation:
$dv_T du_{min} \rightarrow dA'_{max} dT'$. The corresponding Jacobian is
$\mid dv_T/dT' \mid ~ \mid du_{min}/dA'_{max} \mid$
with $\mid dv_T/dT' \mid = R_E/T'^2$. Furthermore we
insert into the above integral:
$\int \delta(A'_{max}- A_{max}) d A_{max}
\int \delta(T'- T) d T =1$ and as next we perform the integration
over $dT'$ and $dA'_{max}$.
This way we get
\begin{equation}
N_{ev} \propto \int d A_{max} d T d\mu \frac{dn_0}{d\mu}
\left| \frac{du_{min}(A_{max})}{d A_{max}} \right|
\frac{\mu^2}{ T^4}
\int (x(1-x))^2 H(x)
exp\left( -\frac{r_E^2 \mu x(1-x)}{v^2_H  T^2} \right) dx~.
\label{eqno:7}
\end{equation}
{}From this expression we can define, up to a normalization constant,
the probability $P$ that a microlensing event of
duration $ T$ and  maximum amplification $A_{max}$
be produced by a MHO of mass $\mu$, that we see first of all
is independent of $A_{max}$
\begin{equation}
P(\mu,T) \propto \frac{\mu^2}{ T^4} \int_0^1 dx (x(1-x))^2 H(x)
exp\left( -\frac{r_E^2 \mu x(1-x)}{v^2_H  T^2} \right) ~. \label{eqno:8}
\end{equation}
We also see that $P(\mu, T)=P(\mu/ T^2)$. The measured values for
$ T$ are listed in table 1.

In figure 1 we plot $P(\mu,T)$ for the event
detected by the MACHO collaboration. (The figures corresponding
to the EROS events have the same shape. One has to multiply the number
on the $\mu$ axis by $(T/16.9~ days)^2$ to get the new figures.)
The normalization
is arbitrarily
chosen such that the maximum of $P(\mu_{MP}, T)=1$, and $\mu_{MP}$ is
the most probable value.
We get that the maximum corresponds to
 $\mu r_E^2/v^2_H T^2=13.0$. This leads to the most probable value
for $\mu$ for this event: $\mu_{MP}=127~(v_H/210~ km/s)^2$.
The corresponding most likely value of $\mu$
for the other two events
are obtained using this relation by inserting the appropriate value for
$T$. From figure 1 we also see that the 50\% confidence interval
embraces for the mass $\mu$ approximately
the range $1/3\mu_{MP}$ up to $3 \mu_{MP}$.
The most probable values for the mass of the three
events calculated with $P(\mu,T)$ are listed in table 1.
The average of the most probable masses of the three events gives
$<\mu_{MP}>=284$ $(v_H/210~km/s)^2$,
corresponding to
0.27 $M_{\odot}$ for $v_H=210~ km/s$.

A more systematic way to extract information on the masses is to use the
method of moments as presented in ref.\cite{kn:Derujula}. The moments
$<\mu^m>$ are given by\footnote{We neglect the efficiency function
$\epsilon_n(\mu)$ \cite{kn:Derujula},
which for the range that is relevant here
is practically identically equal to 1.}
\begin{equation}
<\mu^m>=\int d\mu \frac{dn_0}{d\mu}\mu^m~, \label{eqno:10}
\end{equation}
and $<\mu^m>$ is related to $<\tau^n>=\sum_{events} \tau^n$,
with $\tau \equiv (v_H/r_E) T$, as constructed
from the observations by
\begin{equation}
<\tau^n>=\int dN_{ev} \tau^n=V u_{TH} \Gamma(2-m) \widehat H(m) <\mu^m>~,
\label{eqno:11}
\end{equation}
with $m \equiv (n+1)/2$ and
\begin{equation}
V \equiv 2 N_{\star} t_{obs}~ D~ r_E~ v_H=26~ pc^3~ \frac{N_{\star}}{10^6}
{}~\frac{t_E}{4~ months}~, \label{eqno:12}
\end{equation}
\begin{equation}
\Gamma(2-m) \equiv \int_0^{\infty} \left(\frac{v_T}{v_H}\right)^{1-n}
f(v_T) dv_T~,
\label{eqno:13}
\end{equation}
\begin{equation}
\widehat H(m) \equiv \int_0^1 (x(1-x))^m H(x) dx~.  \label{eqno:14}
\end{equation}

The mean local density of MHO (number per cubic parsec)
is $<\mu^0>$. The average local mass density in MHO is
$<\mu^1>$ Jupiter masses per cubic parsec. The mean MHO mass, which we get from
the three events, is
\begin{equation}
\frac{<\mu^1>}{<\mu^0>}=152~ \left(\frac{v_H}{210~ km/s}\right)^2~,
\label{eqno:aa}
\end{equation}
or 0.144 $M_{\odot}$ for $v_H=210~ km/s$.
(To obtain this result we used the values of $\tau$
as reported in table 1, whereas $\Gamma(1)\widehat H(1)=0.0362$ and
$\Gamma(2)\widehat H(0)=0.280$ as
plotted in figure 6 of ref.\cite{kn:Derujula})
The average value deduced from the most probable mass via $P$ is somewhat
higher than the one deduced using the mass moment method,
but it has of course not to be the same.

Although based on few events, these results are of
interest and show clearly that the three observations lead to
a similar mass range for the MHO. The average values are higher than the
lower limit for nuclear ignition, which is 0.08 $M_{\odot}$.
(Notice that these values increase for values $v_H > 210
{}~ km/s.)$ Nevertheless,
even if it will be confirmed that the dark compact halo objects have
masses in the range 0.1 to 0.3 $M_{\odot}$, their luminosity
would be too small to be detectable with the present telescopes.
It may well be that the next generation of infrared satellites to be put
in orbit like ISO or SIRTF will be able to detect them, either through
measuring their integrated luminosity in halos of nearby galaxies or
by observing objects which are in the solar vicinity \cite{kn:daly}.

Once more data will be available other important quantities
such as the statistical error in (\ref{eqno:aa}), and the fraction
$f \equiv {M_J}/{\rho_0} \sim 0.12~pc^2 <\mu^1>$
of the local
dark mass density (the latter one given by $\rho_0$) detected
in the form of MHOs can be determined. Possible consequences as a result
of such a determination on the problem of dark matter
have been discussed in ref.\cite{kn:Pad,kn:Turner}.\\

Table 1: For the observed three microlensing events (Am = American-Australian
collaboration event; Fr1 and Fr2 are respectively the first and the second
event of the French collaboration) we give here the most probable mass as
obtained by $P(\mu, T)$. With $M_{-}$ and $M_{+}$ (in $M_{\odot}$ units)
we give respectively the lower and upper limit of the 50\% confidence
interval, within which $P(\mu, T) \geq 0.5$.

\begin{center}
\begin{tabular}{|c|c|c|c|}\hline
  & Am & Fr1 & Fr2 \\
\hline
$ T$ (days) & 16.9 & 27 & 30 \\
\hline
$\tau (\equiv \frac{v_H}{r_E} T)$& 3.14 & 5.01 & 5.57 \\
\hline
$\mu_{MP}$ & 127 & 325 & 401 \\
\hline
$M_{MP}/M_{\odot}$ & 0.12 & 0.31 & 0.38 \\
\hline
$M_{-}/M_{\odot}$& 0.04 & 0.11 & 0.13 \\
\hline
$M_{+}/M_{\odot}$ & 0.38 & 0.95 & 1.2 \\
\hline
\end{tabular}
\end{center}

\noindent{\bf Acknowledgments}\\

We would like to thank A. De R\'ujula for very useful
discussions.\\

\newpage
\begin{reflist}
\bibitem{kn:Derujula1} A. De R\'ujula, Ph. Jetzer and
E. Mass\'o, Astron. \& Astrophys. {\bf 254}, 99 (1992).
\bibitem{kn:Pac} B. Paczy\'nski, Astrophys. J. {\bf 304}, 1 (1986).
\bibitem{kn:EROS} E. Aubourg et al., Nature {\bf 365}, 623 (1993).
\bibitem{kn:MACHO} C. Alcock at al., Nature {\bf 365}, 621 (1993).
\bibitem{kn:Udalski} A. Udalski et al., Acta Astron. {\bf 43}, 289 (1993).
\bibitem{kn:Derujula} A. De R\'ujula, Ph. Jetzer and E. Mass\'o,
Mont. Not. R. astr. Soc. {\bf 250}, 348 (1991).
\bibitem{kn:Griest} K. Griest, Astrophys. J. {\bf 366}, 412 (1991).
\bibitem{kn:daly} R.A. Daly and G.C. McLaughlin, Astrophys. J. {\bf 390},
423 (1992).
\bibitem{kn:Pad} T. Padmanabhan and K. Subramanian,
Preprint IUCAA-23, Ganeshkhing, India (1993).
\bibitem{kn:Turner} M. Turner, Fermilab-PUB-93-298-A (1993).
\end{reflist}
\begin{figcap}

\item The function $P(\mu, T)$ for the American-Australian event as
a function of $log~ \mu$. $P(\mu, T)$
is normalized such that it is equal to
1 for the most likely mass value.

\end{figcap}

\end{document}